\PassOptionsToPackage{svgnames}{xcolor}
\PassOptionsToPackage{dvipsnames}{xcolor}
\PassOptionsToPackage{table}{xcolor}

\documentclass[acmsmall]{acmart}

\author{Mootez Saad}
\affiliation{%
  \institution{Dalhousie University}
  \country{Canada}}
\email{mootez@dal.ca}

\author{Tushar Sharma}
\affiliation{%
  \institution{Dalhousie University}
  \country{Canada}}
\email{tushar@dal.ca}

\AtBeginDocument{%
  \providecommand\BibTeX{{%
    \normalfont B\kern-0.5em{\scshape i\kern-0.25em b}\kern-0.8em\TeX}}}

\usepackage{multicol}
\usepackage{listings}
\usepackage{etc}
\usepackage{booktabs}
\usepackage{float}
\usepackage{multirow}
\usepackage{graphicx}
\usepackage{textcomp}
\usepackage{colortbl}
\usepackage{balance}
\usepackage{subfigure}
\usepackage{pifont}
\usepackage{makecell}
\usepackage{url}
\usepackage{algorithm}
\usepackage{algpseudocode}
\usepackage{hyperref}
\usepackage{tabularx}
\usepackage[frozencache=true,cachedir=minted-cache]{minted}
\usepackage[most]{tcolorbox}

\def\BibTeX{{\rm B\kern-.05em{\sc i\kern-.025em b}\kern-.08em
    T\kern-.1667em\lower.7ex\hbox{E}\kern-.125emX}}

\newcommand{\xmark}{\ding{55} }
\newcommand{\concord}{{\sc Concord}}

\newcommand{\abst}{{\sc ast}}
\newcommand{\pdg}{{\sc pdg}}
\newcommand{\cfg}{{\sc cfg}}

\newcommand{\gnn}{\textsc{gnn}}

\newcommand{\mcc}{\textsc{mcc}}

\newcommand{\topic}[1]{\vspace{2mm}\noindent\textbf{#1:}}

\newcommand{\keyword}[1]{\textcolor{blue}{\textbf{#1}}}

\newcommand{\stringliteral}[1]{\textcolor{purple}{'#1'}}
\newcommand{\code}[1]{\texttt{#1}}

\definecolor{codegreen}{rgb}{0,0.6,0}
\definecolor{codegray}{rgb}{0.5,0.5,0.5}
\definecolor{codepurple}{rgb}{0.58,0,0.82}
\definecolor{backcolour}{rgb}{0.96,0.96,0.96}

\definecolor{codebg}{HTML}{F7F7F7}
\definecolor{codefg}{HTML}{5E5E5E}
\definecolor{keyword}{HTML}{0070C0}
\definecolor{string}{HTML}{FF8000}

\newcommand{\dl}{{\sc dl}}
\newcommand{\rone}{{\sc{R$_1$}}}
\newcommand{\rtwo}{{\sc{R$_2$}}}
\newcommand{\rthree}{{\sc{R$_3$}}}

\lstdefinestyle{concord_style}{
	backgroundcolor=\color{backcolour},   
	commentstyle=\color{codegreen},
	keywordstyle=\color{blue},
	stringstyle=\color{codepurple},
	basicstyle=\ttfamily,
	breakatwhitespace=false,         
	breaklines=true,                 
	captionpos=b,                    
	keepspaces=true,                                   
	numbersep=6pt,                  
	showspaces=false,                
	showstringspaces=false,
        numbers=none,
        frame=single,
	showtabs=false,                  
	tabsize=2,
	xleftmargin=.2in,
        morekeywords={Edge, Node, Tasks, Conditions, Representations},
        moredelim=**[is][\color{Fuchsia}]{@}{@},
        moredelim=**[is][\color{OliveGreen}]{|}{|},
        moredelim=**[is][\color{Violet}]{`}{`},
        moredelim=**[is][\color{RedOrange}]{-}{-},
}
\lstdefinestyle{java_style}{
    language=Java, 
    keywordstyle=\bfseries\color{blue}, 
    stringstyle=\color{red}, 
    showstringspaces=false, 
	breakatwhitespace=false,         
	breaklines=true,                 
	captionpos=b,                    
	keepspaces=true,                 
	numbers=left,                    
	numbersep=6pt,                  
	showspaces=false,                
	showtabs=false,                  
	tabsize=2,
	xleftmargin=.2in,
        escapeinside=||
}
\lstdefinestyle{java_style2}{
    language=Java, 
    basicstyle=\ttfamily\footnotesize, 
    keywordstyle=\bfseries\color{blue}, 
    stringstyle=\color{red}, 
    showstringspaces=false, 
	breakatwhitespace=false,         
	breaklines=true,                 
	keepspaces=true,                 
	showspaces=false,                
	showtabs=false,                  
	tabsize=2,
	xleftmargin=.2in,
        escapeinside=||,
        linewidth=\linewidth
}
\lstdefinestyle{python}{
    language=Python,
    basicstyle=\ttfamily\footnotesize,
    keywordstyle=\color{keyword},
    stringstyle=\color{string},
    showstringspaces=false,
    tabsize=4,
    breaklines=true,
    numbers=left,
    numberstyle=\footnotesize\color{codefg},
    captionpos=b,
    numbersep=5pt,
}
\begin{document}

\title{CONCORD: Towards a DSL for Configurable Graph Code Representation}


\begin{abstract}
Deep learning is widely used to uncover hidden patterns in large code corpora. To achieve this, constructing a format that captures the relevant characteristics and features of source code is essential. Graph-based representations have gained attention for their ability to model structural and semantic information. However, existing tools lack flexibility in constructing graphs across different programming languages, limiting their use. Additionally, the output of these tools often lacks interoperability and results in excessively large graphs, making graph-based neural networks training slower and less scalable.

We introduce \concord{}, a domain-specific language to build customizable graph representations. It implements reduction heuristics to reduce graphs' size complexity. We demonstrate its effectiveness in code smell detection as an illustrative use case and show that: first, \concord{} can produce code representations automatically per the specified configuration, and second, our heuristics can achieve comparable performance with significantly reduced size. 
\concord{} will help researchers 
a) create and experiment with customizable graph-based code representations for different software engineering tasks involving DL,
b) reduce the engineering work to generate graph representations,
c) address the issue of scalability in GNN models, and
d) enhance the reproducibility of experiments in research through a standardized approach to code representation and analysis.

\end{abstract}


\maketitle

\section{Introduction}
In recent years, deep learning (\dl{})-based methods for source code analysis have gained significant momentum~\cite{Sharma2021Survey, Allamanis2018a}. 
Such methods have been applied in various software engineering tasks, such as defect prediction~\cite{Li2022SySeVR},~\cite{Grieco2016},~\cite{Zou2021mu}, code smell detection~\cite{Sharma2021A},~\cite{Liu2019CodeSmells},~\cite{Hadj2018hybrid}, code summarization~\cite{Haque2020},~\cite{Iyer2016}, ~\cite{Bansal2021project}, and program synthesis~\cite{Sharma2021Survey, Allamanis2018a}.
Generating code embeddings is a common step in these approaches, where code is converted from its raw textual form to a numeric representation that can be processed and fed to \dl{}-models~\cite{Sharma2021A, feng2020codebert,cubert,Nguyen2016}.
However, treating source code as a stream of tokens ignores its rich structure and semantics~\cite{code2vec},~\cite{DeFreez2018},~\cite{Devlin2017SemanticCR},~\cite{buch2019}. 
Hence, efforts have been made to incorporate the code's structural and semantic features into its vector representation \cite{code2vec, DeFreez2018, Devlin2017SemanticCR, buch2019}.

To further leverage the power of semantics, researchers are experimenting with graph-based neural network architectures~\cite{astnn, Zhou2019a} that can better capture  different semantic aspects (\eg{} control and data flow) of source code.
Often, based on the downstream task,
it is required to combine multiple code representations 
such as Abstract Syntax Tree (\abst{})
and Program Dependence Graph (\pdg{})
into a unified graph structure.
For example,
Allamanis \etal{}~\cite{Allamanis2018} augmented \abst{} with custom \textit{data flow edges} and \textit{control flow edges}
to detect variable misuse in a code snippet.
The process of combining code representations, 
pre-processing them, 
and training a \dl{} model is done manually and from scratch
which wastes significant research time and resources.

Current tools for constructing graph representations of source code face challenges due to different parsers used, resulting in varied \abst{}s representation and hindering effective combinations. Lack of interoperability arises from generating graphs in different formats (JSON, GraphML), and functionality fragmentation occurs due to unique edge augmentation operations introduced by each tool. Combining these representations or artifacts requires considerable effort, as noted by Siow~\etal{}~\cite{Siow2022LearningPS}.

Another significant problem in generating graphs from source code is the high size complexity such structures can attain. 
To demonstrate this issue, let us consider a small code snippet and its corresponding \abst{}:

\begin{figure}[h]
\begin{minipage}{.4\textwidth}
\input{assets/listings/example_c_like}
\end{minipage}
\qquad 
\begin{minipage}{0.5\textwidth} 
    \includegraphics[width=\linewidth]{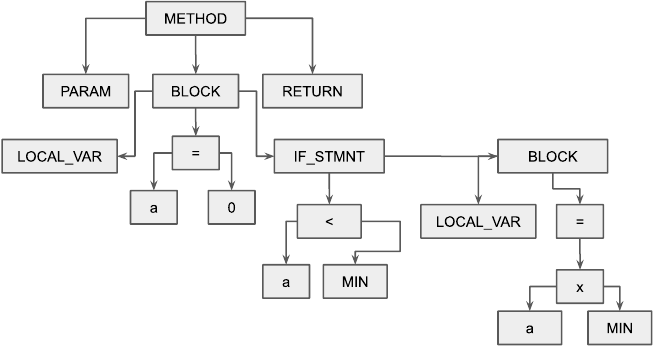}
    \caption{\abst{} of Listing \ref{lst:example_clike}}
    \label{fig:example_abst}

\end{minipage}
\end{figure}
Despite being only four lines long, this code results in an \abst{} containing 19 nodes as shown in Figure~\ref{fig:example_abst}. 

An \abst{} can only capture the structural aspects of code, which is why it might be required to include other relevant attributes to represent control and data flow.
Therefore, in order to effectively model these properties, 
additional structures such as Control Flow Graphs (\cfg{}s) need to be incorporated. 
However, this augmentation further aggravates the issue of complexity in terms of graph size. 

Graph Neural Networks (\gnn{}s) require large memory and computational resources during training, 
which renders them computationally expensive to train and difficult to scale. 
This is because \gnn{}s operate on graph data, which is often more complex and larger than other types of data as we have shown in the previous example.
During training, a \gnn{} must store and manipulate large adjacency matrices representing the graph. 
Additionally, these models typically perform multiple rounds of message passing or graph convolution operations adding further to memory requirements. 
This entails that computational requirements are directly proportional to the size of the graph and the number of layers in a \gnn{}. 
Hence, finding reduction techniques is critical to make the training tractable.

A codebase may contain \textit{noisy} statements that do not contribute to its core logic. Finding unrelated print statements is natural: often, developers use print statements for debugging and tracing errors. Studies have shown~\cite{Beller2018, Liu2023debug} that professional software developers rely on print statements such as \texttt{printf} as a debugging method. These statements do not relate to the overall semantics because debugging is not the primary goal of the said snippet. We can claim the same about temporary variables. We introduce temporary variables to make source code more readable for others, however, this also comes at a cost by making the program longer.
One might conjecture that the removal of such statements can \textit{reduce} source code and allow the model to learn better its implicit patterns and salient features to capture its intended behaviour. 
Moreover, such modifications to the code while generating its representation highly depend on the downstream task. The existing tools and technologies do not offer any customizable way to generate code representation automatically.

This paper introduces {\sc \underline{con}}figurable {\sc \underline{co}}de {\sc \underline{r}}epresentation (\concord{}), a Domain Specific Language (\dsl{}) and a unifying framework designed to automate the generation of custom code representations. The \dsl{} automatically generates code representations from source code and as per the specified configuration. By leveraging \concord{} researchers can easily obtain customized code representations for their downstream tasks without the need to handle intricate code representation details. 
This streamlined approach not only enhances overall efficiency but also reduces the risk of errors. Experimental results demonstrate that our proposed method \textit{achieves improved model performance at a lower computational cost.} We applied \concord{} to solve the code smell detection task, and managed to maintain up to 100\% performance and reduce computations by 10.15\%.
\newline
We summarize the key contributions of this work below.
\begin{itemize}
    \item We provide a one-stop solution in the form of a \dsl{} that unifies multiple edge augmentation techniques from previous works to produce rich and expressive graph representations in a \textit{composable}, \textit{configurable}, and \textit{cross-linguistic} manner.
    \item We propose and implement two pruning techniques to reduce the size of code graphs to make training more tractable.
    \item We present a case study to demonstrate the capabilities of \concord{} to apply on code smell detection task.
    \item We show how the proposed pruning techniques can reduce size and computational time for \gnn{} training while preserving performance.
    \item We have made the \textbf{replication package} available online~\cite{Replication} for other researchers to use and extend the proposed \dsl{} and framework.
\end{itemize}
\section{Background and Related Work}
\subsection{Graph Neural Networks}
Graph Neural Networks~\cite{Scarselli2009} operate on graph-structured data, using a message-passing mechanism to learn node and edge representations. The process involves computing messages, aggregating them, and updating node representations iteratively.
They have been used in the context of Software Engineering to solve different tasks, specifically those that focus on source code analysis. Such works include but are not limited to defect prediction~\cite{Zhou2019a, Chakraborty2022}, \cite{Cui2022R},~\cite{Liu2023},~\cite{Li2022}, clone detection~\cite{Wang2020DetectingCC},~\cite{Mehrotra2021}, code search~\cite{Liu2021gsn},~\cite{Zeng2023},~\cite{Wan2020}, and code summarization~\cite{LeClair2020}, \cite{Liu2020RetrievalAugmentedGF}.
The reason behind the prevalence of using such architecture is that source code is often modelled using graphs like in the case of syntax trees, control flow graphs and call graphs. Such structures encompass properties that cannot be reflected if the source code was represented as a flat sequence of tokens.
\subsection{Domain Specific Languages}
A domain-specific language (\dsl{}) is a specialized language that uses concepts and rules of a domain \cite{fowler}. 
\dsl{}s are divided into two major categories---\textit{external} and \textit{internal}. 
External \dsl{}s like {\sc css} and {\sc sql}, have their own syntax and parser that interprets the language or translates it into another language.
Internal \dsl{}s are written on top of an existing general-purpose language such as libraries or {\sc api}s. 
The proposed \dsl{} \concord{} falls into the first category. 
There have been many \dsl{}s proposed in software engineering literature~\cite{Thanhofer2017}.
For example,
Pig Latin~\cite{Olston2008PigLA} is a textual \dsl{} that abstracts the MapReduce implementation. 
The \dsl{} code is translated by the compiler into MapReduce jobs that are executed by the infrastructure provided by Apache Hadoop. 
Andrzejak \etal{} proposed {\sc nldsl}~\cite{Andrzejak2019} that is used to create chain and pipeline operations for data processing. Giner-Miguelez \etal{}~\cite{GinerMiguelez2023} presented a \dsl{} to precisely describe machine learning datasets, aligning with the data-centric cultural shift in the {\sc ML} community. The \dsl{} aids in dataset selection and result reproducibility, with a Visual Studio Code plugin for usability. Case studies and experiments validate its expressiveness and potential impact on achieving higher-quality {\sc ML} models, particularly in terms of fairness, diversity, and bias mitigation.
Morales \etal{}~\cite{Morales2022} have proposed a \dsl{} to model {\sc AI} engineering processes, tailored for multidisciplinary teams developing {\sc AI}-embedded software. It provides a structured framework and common ground to address communication issues and promote best practices. Its goal is to facilitate the formalization of {\sc AI} processes within organizations and seamlessly integrate with existing model-driven tools, advancing the adoption of AI engineering practices.

\begin{figure*}
    \centering
    \includegraphics[width=\columnwidth]{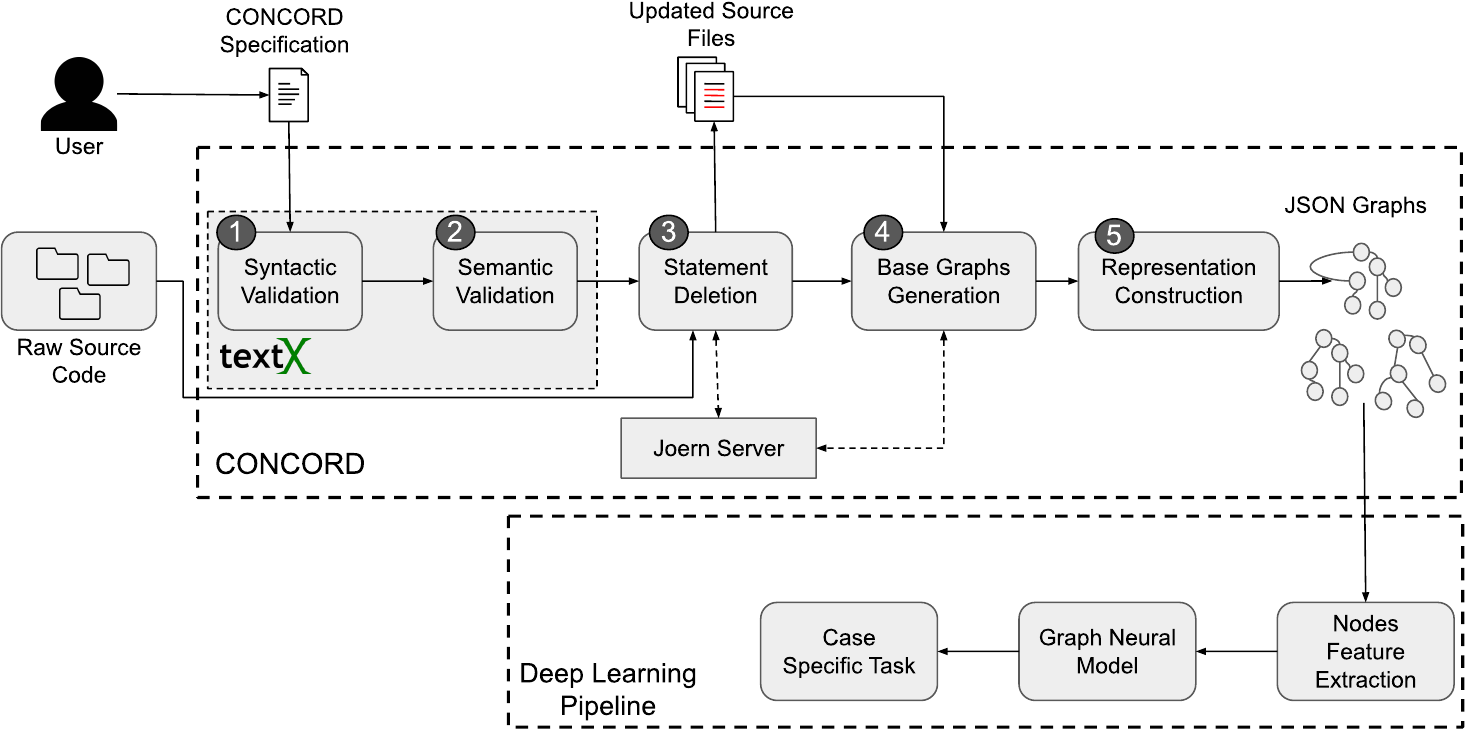}
    \caption{Overview of CONCORD}
    \label{fig:concord_overview}
\end{figure*}

\subsection{Tools for code graphs construction}
Bieber \etal{}~\cite{Bieber2022} provide an open-source Python library called \textit{python\_graphs} 
to construct graph representations of Python programs suitable for training machine learning models by employing static analysis.
The library is capable of constructing control-flow graphs, data-flow graphs, and composite ``program graphs'' that combine control-flow, data-flow, syntactic, and lexical information about a program. 
Rice \etal{}~\cite{Rice2018} developed a \textit{javac} plugin \textit{feature-javac} that creates feature graphs for Java-based source code repositories. 
These graphs incorporate a variety of syntactic and semantic features. 
It bears similarities with the \textit{python\_graphs} library and has been designed to facilitate machine learning on source code research.

\textbf{Limitations of existing work:}
The first limitation is that each tool supports one programming language, either Python or Java, thereby limiting their applicability to those languages only. 
Second is the lack of \textit{composability} and \textit{configurability}. These tools produce representations that include all types of edge augmentations at once. In other words, users cannot specify configurations that specify what augmentations to be included or excluded. Finally, the set of augmentations in these tools is disjoint, \ie{} a list of operations is found in one tool but missing in another. This fragmented nature of functionalities makes it difficult to inter-operate the structures yielded by these methods.

\subsection{Source code reduction}
Zhang \etal{}~\cite{Zhang2022} introduce an approach referred to as \textit{DietCode} that leverages large pre-trained models such as CodeBERT~\cite{feng2020codebert} for source code. 
Through an empirical analysis of CodeBERT's attention mechanism, 
they discovered that the model attends more to certain types of tokens and statements such as keywords and data-relevant statements. 
Based on this finding, they proposed three strategies for simplifying CodeBERT's input program: 
\textit{word dropout}, 
\textit{frequency filtering}, 
and an attention-based approach that \textit{selects statements and tokens} with the highest attention weights. 
They tested the efficacy of \textit{DietCode} on code search and code summarization tasks and found that it performs comparably to CodeBERT while requiring 40\% less computational cost during fine-tuning and testing.
Rabin \etal{}~\cite{Rabin2021} propose \textit{sivand}, 
which employs delta debugging to identify the most critical code segments for a deep learning model. 
This technique involves a deep learning model receiving a code snippet as input, which is subsequently segmented. 
The neural network model then processes each segment for a downstream task, such as method name prediction. 
If a segment achieves a desirable score, it is further divided. 
This iterative process continues until the subset's performance fails to meet the targeted score. 
Ultimately, the algorithm produces the smallest code snippet that satisfies the model's objective.

\textbf{Limitations of existing work:}
The limitation of these works is that they require multiple rounds of expensive computation. Specifically, \textit{sivand} requires the execution and evaluation of a deep learning model at each stage of program reduction, which can be inefficient if the data scales upwards. Similarly, at least one forward and backward pass (\ie{} fine-tuning) is needed by \textit{DietCode} to compute the attention scores of each token. This overhead operation is computationally demanding given that the number of parameters ($125$M and $220$M) in the considered Language Models. 
\newline
As we have discussed, graph neural networks have the capacity to better model source code. However, the current state of tooling that addresses the challenges such models bring is limited. The aim is to propose a new solution in the form of a domain-specific language that enables flexible and configurable graph code modeling to address such limitations.
\section{Concord DSL Design}
\begin{figure}[H]
    \centering
    \includegraphics[scale=.5]{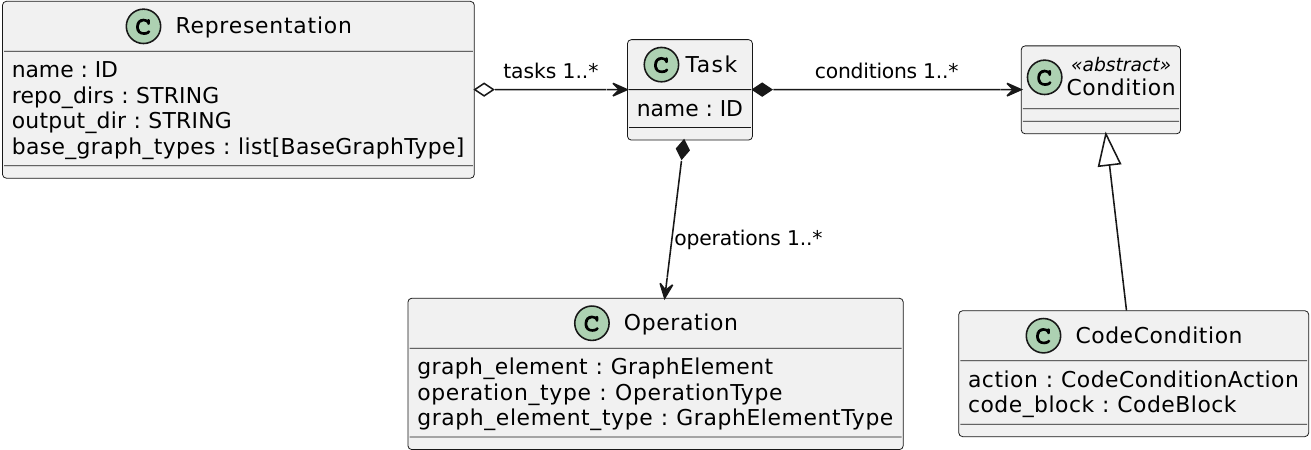}
    \caption{CONCORD's metamodel excerpt}
    \label{fig:concord_metamodel}
\end{figure}

\concord{} is a \dsl{} and framework that we propose in this paper to construct customized graph representations of source code.
The constructed graph representation can be employed for training \dl{} models. 
Figure~\ref{fig:concord_overview} provides an overview of a generic pipeline 
for \dl{} model training 
using \concord{}. 
To train a \gnn{} model, a user first generates a graph representation of the source code. They provide a \concord{} configuration which is interpreted, and the language's backend generates the concrete samples. These samples can later be used for training and inference.
The language grammar is defined and interpreted using \textit{textX}~\cite{Dejanovic2017}. A \dsl{} is typically characterized by a metamodel, which serves as a representation of its domain-specific entities and their interconnections. Figure~\ref{fig:concord_metamodel} illustrates \concord{}'s metamodel and the relationship between the \dsl{}'s elements.

\concord{}'s grammar is formalized by a Parsing Expression Grammar (PEG)~\cite{Ford2004}. PEG is a formalism used to describe language syntax and facilitate  parsing. Unlike Context-Free Grammar (CFG), PEG ensures deterministic parsing and disallows ambiguity. It employs recursive descent parsing, prioritizes alternatives in order, handles left recursion gracefully, and provides better support for nested structures. In Table~\ref{tab:concord_grammar_rules} we present the different grammar rules and their body. The full grammar specification can be found in the replication package.\footnote{\url{https://bit.ly/3Kb5u84}}

\begin{table}[H]
    \centering
    \caption{Grammar Rules for CONCORD}
    \resizebox{.75\columnwidth}{!}{%
    \begin{tabular}{ll}
    \toprule
    \textbf{Rule} & \textbf{Rule body} \\
    \midrule
       \rowcolor{lightgray!25} BaseGraphType & AST$|$CFG$|$PDG \\ 
        CodeBlock & catch$|$for$|$while$|$if$|$else \\ 
       \rowcolor{lightgray!25} CodeConditionAction & exclude$|$include \\ 
        Comment & $\backslash$/$\backslash$*(.|$\backslash$n)*?$\backslash$*$\backslash$/|$\backslash$/$\backslash$/.*?\$ \\ 
       \rowcolor{lightgray!25} EdgeType & \makecell{next\_token$|$next\_sibling$|$for\_cfg$|$while\_cfg$|$last\_read\_write$|$\\guarded\_by$|$returns\_to$|$computed\_from$|$last\_lexical\_use} \\ 
        GraphElement & Node$|$Edge \\ 
       \rowcolor{lightgray!25} NodeType & print$|$logging$|$sys\_exit$|$simple\_assignment \\ 
        OperationType & add$|$remove \\ \bottomrule
    \end{tabular}}
\label{tab:concord_grammar_rules}
\end{table}

\subsection{DSL key elements}
A \textit{Representation} specifies the intended graph to be constructed;
it is defined by a set of base representations, as well as a set of \textit{Tasks}. 
The base representations are: \abst{}, \cfg{}, and \pdg{}. 
If multiple base representations are specified, they are merged into a single graph.

A \textit{Task} comprises a set of \textit{Operations} that are applied to the base graphs. 
An \textit{Operation} represents an atomic action that can be applied to the graph, 
and there are two types---\textit{node removal} and \textit{edge addition}. 

A \textit{Code Condition} specifies the code structures that are exempt from node removal. 
Currently, supported code structures include control statements such as \textit{if} statements and \textit{for} loops (rule \textit{CodeBlock} in Table~\ref{tab:concord_grammar_rules}). 
In other words, nodes (\ie{} statements) that are within these blocks will not be removed, if specified. 
This feature 
provide greater flexibility to the user, where instead of blindly pruning statements through all code regions, they may want to keep some blocks intact depending on their use case.

\subsection{Graph generation process}
\noindent
\textbf{Syntactic validation:}
In the syntactic evaluation, \textit{textX}~\cite{Dejanovic2017} constructs a model from the \concord{} configuration file and checks whether it conforms to the grammar rules and the metamodel of the language.
\noindent
\newline
\noindent
\textbf{Semantic validation:}
We ensure the correctness of representation semantics. 
For example, \texttt{Node remove next\_token} is syntactically correct, but it is semantically invalid since \textit{next\_token} is an edge addition operation.
The framework warns the user when it detects such issues.
\newline
\noindent
\textbf{Statement deletion:} In this step, we prune the statements specified by the user at the file level. 
\newline
\noindent
\textbf{Base graphs generation:}
Once the configuration passes the validation steps and statements are pruned,
we generate the base graph representations by communicating with \textit{joern}~\cite{Yamaguchi2014} (a platform for code analysis that supports C/C++, Java, Javascript, Python and PHP), through its HTTP server.
The generated representations are stored in {\sc json} format.
\newline
\noindent
\textbf{Representation construction:}
Finally, we import the base graphs and construct the representation using the \textit{NetworkX} library~\cite{SciPyProceedings_11}. 
The framework generates code representations as per the specification in the configuration file.
We store code representations in {\sc json} format, which can be parsed by NetworkX, allowing further processing, if needed.

\subsection{Node removal}
\label{node_removal}
When source code is transformed into graphs, 
the resulting structure can be significantly complex in terms of size (\ie{} the number of vertices and edges).
To mitigate this issue, one approach is to discard graphs from the dataset that exceed a certain threshold. 
For example, Zhou \etal{}~\cite{Zhou2019a} do not consider graphs with more than $500$ nodes. 
However, such a strategy can lead to information loss, which may adversely impact the performance of \dl{} models. 

Gu \etal{}~\cite{Gu2021} introduced \textit{Simplified Syntax Tree} (\textsc{sst}), 
which is an approximation that simplifies the \abst{} by pruning nodes such as type declarations and modifier keywords. 
By using this simplified representation, they were able to improve the performance of neural models in code search tasks using the CodeSearchNet benchmark. 
Inspired by this heuristic, we introduce a similar way of simplification in \concord{}, however, at a wider (\ie{} statement-level removal) scale. 
The three types of statements that can be removed are \textit{simple assignment}, \textit{print}, and \textit{system exit} statements, 
as a means to reduce the complexity of the graph representations 
(Step $3$, Figure~\ref{fig:concord_overview}).

Simple assignment statements initialize a variable with a literal value, such as an integer, float, double, or boolean value.
Print statements are functions or methods provided by the programming language to output information on console, 
such as the \texttt{println()} statement in Java. Similarly, exit statements are methods used to transfer control of the program, such as the \texttt{System.exit()}. 
In our approach, we perform statement removal at the file level
\ie{} we identify the line numbers corresponding to these statements and then perform the removal for each source code file in a repository. 
We chose to perform statement removal at the file level rather than removing them after generating the base graphs in order to leverage the data flow engine of the \textit{joern} tool for generating the \pdg{}. 
This is especially relevant when the user specifies more complex base representations, such as \pdg{} or \cfg{}, 
as it ensures more reliable results by using a well-implemented and maintained engine.

\begin{figure}[H]
    \centering
    \includegraphics[width=.7\columnwidth]{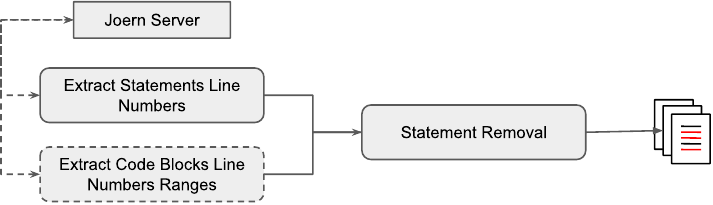}
    \caption{Statement removal process using CONCORD}
    \label{fig:statement_removal_process}
\end{figure}

The overall process is illustrated in Figure~\ref{fig:statement_removal_process}, where we first collect the statements specified by the user, and then perform the deletion, taking into consideration the code blocks specified in the specification.

Algorithm \ref{algo:simple_assignment_algorithm}
presents the procedure for identifying simple assignment statements. 
After loading the source code into \textit{joern}, 
we extract the \abst{} of each method. 
Then, for each tree, we iterate over its nodes and return those that represent simple assignment statements. 
A statement is considered a simple assignment statement if it satisfies the following four conditions:

\begin{enumerate}
\item The node representing the statement is of type \texttt{ASSIGNMENT}.
\item Its left subtree of the node have only one node of type \texttt{IDENTIFIER}.
\item The non-leaf nodes of its right subtree represents one of the allowed operators (arithmetic, logical, relational, and bitwise).
\item All leaf nodes of its right subtree are literals.
\end{enumerate}

\setlength{\textfloatsep}{2pt}
\begin{algorithm}[H]
    \small
    \caption{Check if an \abst{} node is simple assignment}
    \label{algo:simple_assignment_algorithm}
    
    \begin{algorithmic}[1]
    \Function{isSimpleAssignment}{root}
        \Comment{Returns whether an assignment node is simple}
        \State condition$_1$ $\gets$ root.type == ``ASSIGNMENT"
        \State condition$_2$ $\gets$ root.children.\Call{Count}{isIdentifier} == 1
        \State nonLeafNodes $\gets$ \Call{GetNonLeafNodes}{root}
        \State condition$_3$ $\gets$ nonLeafNodes.\Call{Count}{!isAllowedOperation} == 0
        \State leafNodes $\gets$ \Call{GetLeafNodes}{root}
        \State condition$_4$ $\gets$ leafNodes.\Call{Counter}{isLiteral} == leafNodes.size()
        \State \Return condition$_1$ \&\& condition$_2$ \&\& condition$_3$ \&\& condition$_4$
    \EndFunction
    \Statex

            \Function{isAllowedOperation}{node}
        \Comment{Returns whether a node represents an allowed operation}
        \State arithmeticOperation $\gets$ \{ add, sub, mult, div, exp, mod \}
        \State bitwiseOperation $\gets$ \{ not, and, xor, or, shiftLeft, shiftRight \}
        \State logicalOperation $\gets$ \{ not, and, or \}
        \State relationalOperation $\gets$  \{equals, notEquals, greaterThan, lessThan, greaterEqualsThan, lessEqualsThan\}
        \State allowedOperations $\gets$ arithmeticOperation + bitwiseOperation + logicalOperation + relationalOperation
        \State \Return node.type $\in$ allowedOperations
    \EndFunction
    \Statex
    \end{algorithmic}
\end{algorithm}

Although it is sufficient to check whether the right subtree is composed of only one node of type \texttt{IDENTIFIER}, 
we implemented it this way to ensure that even complex statements involving operations between literals, such as \texttt{\textbf{\textcolor{blue}{int}} \textbf{a} \textbf{\textcolor{ForestGreen}{=}} \textbf{\textcolor{Mulberry}{1*7+(1-7)}}}, 
are also considered. 
Though such type of statements are rarely found in practice,
we did so for the sake of completeness.

When removing simple assignment statements we do not remove the entire source code line,
rather, we replace the statement at that specific line with an empty string. 
To explain why we follow such an approach, 
let us consider the example given in Listing~\ref{lst:simple1}.

\begin{listing}[H]
\begin{minted}[
fontsize=\footnotesize,
escapeinside=||
]{cpp}
public static void foo(String[] args) {
   for (|\colorbox{red!25}{int i = 1}|; i <= n; ++i) {
     System.out.println("Printer");
   }
 }
\end{minted}
\caption{Simple assignment statement to be removed}
\label{lst:simple1}
\end{listing}
If we were to remove the entire line, the resulting code would be completely erroneous, as seen in Listing~\ref{lst:simple2}, with mismatched opening and closing brackets.
    


\begin{listing}[H]
\begin{minted}[
fontsize=\footnotesize,
escapeinside=||
]{cpp}
public static void foo(String[] args) {
    
     System.out.println("Printer");
   }
 }
\end{minted}
\caption{Incorrect removal of the statement if the entire line
is deleted}
\label{lst:simple2}
\end{listing}
 However, replacing the statement with an empty string instead preserves the rest of the code that has not matched the condition, which is relevant for the code representation process. 

\begin{listing}[H]
\begin{minted}[
fontsize=\footnotesize,
escapeinside=||
]{cpp}
public static void foo(String[] args) {
   for (; i <= n; ++i) {
     System.out.println("Printer");
   }
 }
\end{minted}
\caption{How the statement is removed by \concord{}}
\label{lst:simple3}
\end{listing}
Moreover, \textit{joern} employs fuzzy parsing~\cite{Koppler1997} to handle incomplete source code with errors, but the accuracy of the output depends on the extent of the error in the input. This approach ensures a careful deletion process that retains code fragments without a significant loss of information.

\subsection{Edge addition}
\label{edge_addition}
The next major operation provided by \concord{} is edge addition. 
Edges generated from base representations,
such as \abst{}s and \pdg{}s,
capture certain aspects of the source code. 
For instance, while \abst{}s effectively reflect the structure of a code snippet, 
they do not capture its control flow. 
As an alternative approach, 
several studies~\cite{Zhou2019a, Allamanis2018, Wenhan2020} have proposed additional sets of edges that can augment the base representations for various tasks. 
In the rest of the sections, we will provide a detailed description of each type of edge that can be added.

\subsubsection{{\sc{NextSibling}}}
Used by Wenhan \etal{}~\cite{Wenhan2020}, this type of edge connects an \abst{} node to its sibling node. 
The reason for such augmentation is primarily due to the behaviour of \textsc{gnn}s that does not consider the order of nodes. 
Hence, it is helpful for the neural model to provide the order of children.

\subsubsection{{\sc{NextToken}}}
\begin{figure}[H]
    \centering
    \includegraphics[width=.3\columnwidth]{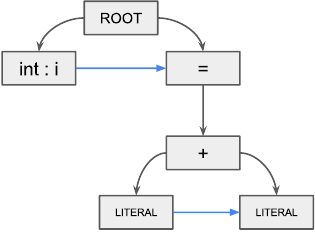}
    \caption{Example of a {\sc{NextToken}} edges in an \abst{}}
    \label{fig:next_token}
\end{figure}

This edge connects terminal nodes of an \abst{} with each other. 
Allamanis \etal{}~\cite{Allamanis2018} introduced this edge type in their study; which was also used by Wenhan \etal{}~\cite{Wenhan2020}. 
The rationale for introducing the edge is that an \abst{} does not reflect the sequential nature of the source code. Let us consider Figure~\ref{fig:next_token} that represents the syntax tree of this C-like statement 
\texttt{\textbf{\textcolor{blue}{int}} \textbf{i} \textbf{\textcolor{ForestGreen}{=}} \textbf{\textcolor{Mulberry}{1+1}}}

\noindent
The figure shows that the {\sc{NextToken}} edges in blue reflect,
to an extent,
the sequential property of the code snippet.

\subsubsection{{\sc{LastRead}}}
Let \texttt{U} is to the set of tokens where a variable \texttt{var} could have been last used. 
This set can consist of multiple nodes, such as when the variable is used in both paths of a conditional statement, or even tokens that come after in the code, as in the case of loops~\cite{Allamanis2018}. 
The {\sc{LastRead}} edges create a connection between these nodes in the set \texttt{U} and \texttt{var}.

\subsubsection{{\sc{LastWrite}}~\textup{\cite{Allamanis2018}}}
It is similar to {\sc{LastRead}}, except the set \texttt{U} would contain the tokens where \texttt{var} could have been last written to.

\subsubsection{{\sc{LastLexicalUse}}~\textup{\cite{Allamanis2018}}}
It is also similar to both of the aforementioned edges, however, it is flow independent. 
In other words, all occurrences of variable \texttt{var} in a source code snippet are chained or linked together by this type of edge.

\subsubsection{{\sc{ComputedFrom}}~\textup{\cite{Allamanis2018}}}
This concerns assignment statements. It connects tokens that are on the right-hand side of an assignment statement to the variable on the left-hand side. 
For example, in Figure~\ref{fig:next_token} both of the literal leaf nodes will be connected to the variable \texttt{i} by this edge.

\subsubsection{{\sc{ReturnsTo}}~\textup{\cite{Allamanis2018}}}
Connects a method declaration node in the \abst{} with the node that refers to the method's \texttt{return} statement.

\subsubsection{{\sc{GuardedBy}} and {\sc{GuardedByNegation}}~\textup{\cite{Allamanis2018}}}
\begin{figure}[ht]
    \centering
    \includegraphics[width=.5\columnwidth]{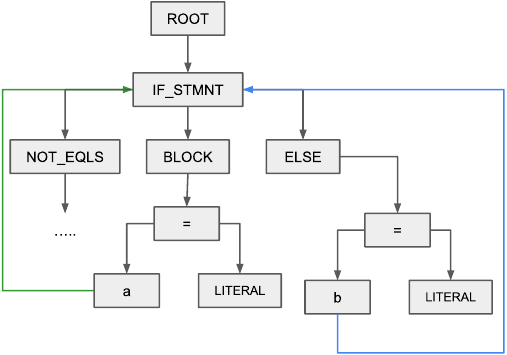}
    \caption{Example of {\sc{GuardedBy}} and {\sc{GuardedByNegation}} edges in an \abst{}}
    \label{fig:guarded_by_neg}
\end{figure}

A {\sc{GuardedBy}} edge connects a conditional node to the usage of the variable when 
the guard condition is true.
{\sc{GuardedByNegation}} performs a similar action but when the conditional is false. 
Figure~\ref{fig:guarded_by_neg} presents the \abst{} of the snippet given below (note that we pruned some parts of it for presentation purposes).
\texttt{
\keyword{if} (\code{a} \stringliteral{!=} \code{b}) \{\code{a} \stringliteral{=} \code{5};\} \keyword{else} \{\code{b}\stringliteral{=}\code{5};\}
}

\noindent
The green edge refers to a {\sc{GuardedBy}} relation, whereas {\sc{GuardedByNegation}} is represented by the blue edge.

\subsubsection{{\sc{WhileCFG}} and {\sc{ForCFG}}}
Both of these relations were introduced by Wenhan \etal{}~\cite{Wenhan2020} to model the control flow behaviour of while and for loops. 
Concretely, each relation is represented by two edges: \texttt{WhileExec} and \texttt{WhileNext} for {\sc{WhileCFG}} and the same logic for {\sc{ForCFG}}. 
The \texttt{WhileExec} connects the conditional node of the while statement to its block node, whereas \texttt{WhileNext} does the inverse. 
Again, the same mechanism is implemented for the for-loop blocks.

Through the implementation of all of these operations, \concord{} provides a unifying tool bench for code graph constructions to deal with the address the current fragmented nature of tools for building such structures.
\section{Case Study: Code Smell Detection}
To demonstrate how \concord{} can be used effectively, we conducted a study to prepare code representations to classify code smells. 
Code smells are symptoms of sub-optimal design and implementation choices manifested in the form of poorly written code that requires refactoring~\cite{Fowler1999}. 
Their presence in a given software system hinders its quality,
specifically maintainability~\cite{Mikhail2011, Gabriele2012} and reliability~\cite{Ikama2022, Foutse2012, Tracy2014, Fehmi2013}.
Hence, many efforts have been made to produce tools and methods to detect them~\cite{Sharma2018}. 
However, an important aspect of code smells is that they are subjective in nature~\cite{Sharma2018, Nandani2023}. 
A code snippet can be smelly in one context, and benign in another. 
This means that classifying such anomalies deterministically using rule-based tools may not produce accurate results. 
To this end, we conduct a set of experiments where we showcase how the features of \concord{} can be put into use in the task of code smell detection. 

In this study, we consider four types of code smells that were initially investigated by Sharma \etal{}~\cite{Sharma2021A}: \textit{Complex method}, \textit{Complex conditional}, \textit{Feature envy}, and \textit{Multifaceted abstraction}. 
We formulate the problem as a one-vs-all binary classification problem, meaning that each code smell has its own dataset where each sample can be either categorized as smelly or non-smelly, which is the commonly adopted way in related literature.

\subsection{Data preparation}
Sharma \etal{}~\cite{Sharma2021A} carried out an empirical study to explore the feasibility to detect code smells using deep learning techniques without feature engineering.
They identified a set of $922$ Java repositories after applying a filtering criteria using RepoRepears'~\cite{Munaiah2017} eight repository characteristics.
We randomly selected and cloned a set of $100$ projects from the repositories used by Sharma \etal{}~\cite{Sharma2021A}. 

To ensure greater reliability,
we selected the snapshot of each project before the replication package by Sharma \etal{}~\cite{Sharma2021A} was published. 
Specifically, we checked out the commit prior to January $26$, $2019$, to reduce the possibility of refactoring that could have led to the removal of code smells.

To demonstrate the capabilities of \concord{} and evaluate the pruning strategies mentioned in Sections~\ref{edge_addition} and~\ref{node_removal}, we define three \concord{} configurations: \rone, \rtwo, and \rthree{} specified in Table~\ref{tab:representation_table}. In Listing~\ref{lst:r2} we show how \rtwo{} is expressed using the \concord{} syntax. All of the specification files for these representations can be found in the replication package\footnote{\url{https://bit.ly/3Q9LpTo}}.
\begin{listing}[h]
\begin{minted}[
fontsize=\footnotesize,
escapeinside=||
]{sql}
Tasks {
    task2 {
        |\textcolor{blue}{Edge}| add next_token
        |\textcolor{blue}{Edge}| add for_cfg
        |\textcolor{blue}{Edge}| add while_cfg
        |\textcolor{blue}{Edge}| add computed_from
        |\textcolor{blue}{Edge}| add guarded_by
        |\textcolor{blue}{Edge}| remove simple_assignment
        Conditions {
            exclude while_block
            exclude if_block
}}}
Representations {
    r2 {
        "/dir/repos_list.csv"
        "output_dir"
        AST
        task2
}}
\end{minted}
\caption{{\rtwo} specification using \concord{}'s syntax.}
\label{lst:r2}
\end{listing}

We select these configurations to capture different aspects of the source code using the \dsl{}. 
Specifically, we utilize \abst{} as the base representation to capture the structural properties of the code for the three representations. 
The {\sc{NextToken}} edge is used to model its sequential nature, while the {\sc{WhileCFG}}, {\sc{ForCFG}}, {\sc{GuardedBy}}, and {\sc{ComputedFrom}} edges are specified to capture the flow of control and data through the program. 
\begin{table}[H]
    \centering
    \caption{Specification of each representation.}
    \resizebox{1\columnwidth}{!}{%
    \begin{tabular}{lcccc}
    \multicolumn{5}{l}{\makecell{
     Nodes: \checkmark means that such statement type (\ie{} set of nodes) is kept, whereas, \xmark signifies that it is removed.\\
    Edges: {\sc{NextToken}} (NT); {\sc{WhileCFG}} (WCFG); {\sc{ForCFG}} (FCG); \\
    {\sc{GuardedBy}} (GB); {\sc{ComputedFrom} (CF);}}} \\
        \toprule
        Representation & \makecell{Base \\ Representation} & \makecell{Print\\Statements} & \makecell{Simple Assignment \\ Statements} & Edges \\
        \midrule
        \rone{} & AST & \checkmark & \checkmark & \makecell{NT, WCFG, FCFG, GB, CF} \\
       \rowcolor{lightgray!25} \rtwo{} & AST & \checkmark & \xmark & \makecell{NT, WCFG, FCFG, GB, CF} \\
        \rthree{} & AST & \xmark & \checkmark & \makecell{NT, WCFG, FCFG, GB, CF} \\
        \bottomrule
    \end{tabular}}
    \label{tab:representation_table}
\end{table}
The representations \rtwo{} and \rthree{} differ from \rone{} as they specify a pruning strategy.

After executing \concord{} using each of the three configurations, 
we generate an initial collection of $\approx$ $1.5M$ methods, 
with $\approx$ $500K$ samples for each configuration. We then filter out methods generated by {\rone} that did not contain at least one print and a simple assignment statement.
To construct the datasets for each code smell, we randomly select a subset of the methods, 
while maintaining the positive to negative ratio mentioned in the work of Sharma \etal{}~\cite{Sharma2021A}. 
Finally, we perform an 80:10:10 split corresponding to training, validation, and test using stratified sampling to ensure fair evaluation for each classifier. Table~\ref{tab:smell_datasets_stats} presents a summary of the training dataset with the class distribution for each dataset at each split.
\begin{table}[H]
\centering
\caption{Statistics of datasets}
\label{tab:smell_datasets_stats}
\resizebox{.65\columnwidth}{!}{%
\begin{tabular}
{@{}lcccccc}
\toprule
\multirow{2}{*}{Smell} & \multicolumn{2}{c}{Train} & \multicolumn{2}{c}{Validation} & \multicolumn{2}{c}{Test} \\
\cmidrule(lr){2-3} \cmidrule(lr){4-5} \cmidrule(lr){6-7}
& N & P & N & P & N & P \\
\midrule
\rowcolor{lightgray!25} Complex Method & 60,000 & 3,222 & 7,500 & 403 & 7,500 & 403 \\
Complex Conditional & 60,000 & 2,478 & 7,500 & 310 & 7,500 & 310 \\
\rowcolor{lightgray!25} Feature Envy & 30,905 & 326 & 3,864 & 41 & 3,864 & 41 \\
Multifaceted Abstraction & 5,184 & 180 & 648 & 23 & 648 & 23 \\
\bottomrule
\end{tabular}%
}
\end{table}
Table~\ref{tab:concord_dataset} tabulates the average number of nodes and edges in each dataset after running \concord{} using the three mentioned configurations.
\begin{table}[H]
\centering
\caption{Average number of nodes and edges for each dataset under each representation}
\resizebox{.75\columnwidth}{!}{%
\begin{tabular}{lcccccc}
\toprule
\multirow{4}{*}{Smell} & \multicolumn{2}{c}{\rone} & \multicolumn{2}{c}{\rtwo} & \multicolumn{2}{c}{\rthree} \\
\cmidrule(r){2-3} \cmidrule(lr){4-5} \cmidrule(l){6-7}
 & \multicolumn{1}{c}{Avg.} & \multicolumn{1}{c}{Avg.} & \multicolumn{1}{c}{Avg.} & \multicolumn{1}{c}{Avg.} & \multicolumn{1}{c}{Avg.} & \multicolumn{1}{c}{Avg.} \\
 & \multicolumn{1}{c}{Nodes} & \multicolumn{1}{c}{Edges} & \multicolumn{1}{c}{Nodes} & \multicolumn{1}{c}{Edges} & \multicolumn{1}{c}{Nodes} & \multicolumn{1}{c}{Edges} \\
\midrule
\rowcolor{lightgray!25} Complex Method & 40.91 & 65.68 & 33.1 & 52.63 & 33.59 & 53.59 \\
Complex Conditional & 40.82 & 65.48 & 32.77 & 52.02 & 33.35 & 53.16 \\
\rowcolor{lightgray!25} Feature Envy & 274.68 & 431.18 & 267.3 & 418.04 & 270.28 & 423.99 \\
Multifaceted Abstraction & 487.28 & 770.42 & 415.0 & 649.12 & 434.6 & 683.49 \\
\bottomrule
\end{tabular}
}
\label{tab:concord_dataset}
\end{table}
One may observe the difference in the number of nodes and edges between the implementation smells (\textit{Complex Conditional/Method}) and design smells (\textit{Multifaceted Abstraction} and \textit{Feature Envy}). 
This arises because we represent the graph of a design smell instance ({\ie} a class) by combining the graphs of its methods. 
Consequently, on average, design smells have a higher number of nodes compared to its implementation counterpart.

During the construction of the datasets, we ensured the presence of a mapping with the aid of the original dataset provided by Sharma \etal{}~\cite{Sharma2021A}. In particular, for each code smell dataset, a CSV file was used to represent the mapping between each code snippet (\ie{} method) and its corresponding graph representations generated using {\rone}, {\rtwo}, and {\rthree}. 
\begin{table}[H]
  \centering
  \caption{Example of how datasets are stored}
  \label{tab:example_csv}
  \resizebox{\columnwidth}{!}{%
  \begin{tabular}{llllllll}
    \toprule
    concord\_id & project & baseline\_file & r1\_file & r2\_file & r3\_file & label & split \\
    \midrule
   \rowcolor{lightgray!25} 23731 & ganttproject & hasCdata.code & hasCdata\_21.json & hasCdata\_28.json & hasCdata\_24.json & 0 & train \\
   421 & greenmail & clear.code &	clear\_42.json & clear\_58.json & clear\_23.json & 0 & test \\
   \rowcolor{lightgray!25} \ldots & \ldots & \ldots & \ldots & \ldots & \ldots & \ldots & \ldots \\
    10806&c5&clear.code &clear\_108.json&clear\_107.json&clear\_104.json&0&val \\
    \bottomrule
  \end{tabular}
  }
\end{table}
Furthermore, the CSV file includes additional columns to indicate the type of split (train, validation, or test). This approach was adopted to ensure that the models are trained and evaluated on the same data, thereby ensuring a fair evaluation. In Table~\ref{tab:example_csv} we show a snapshot on how these CSV files are stored.

\subsection{Model description}
We employ a Gated Graph Neural Network (\textsc{ggnn})-based~\cite{Li_ggnn2016} classifier to solve the code smell classification task. 
We select this architecture over other variants such as Graph Convolutional Networks (\textsc{gcn})~\cite{Kipf2017} or Graph Attention Networks (\textsc{gat})~\cite{Velickovic2018}, 
due to its superior performance demonstrated in other software engineering tasks~\cite{Siow2022LearningPS} including clone detection, code classification, and vulnerability detection.
\begin{figure}[H]
    \centering
    \caption{Gated Graph Neural Network-based classifier}
    \includegraphics[width=\linewidth]{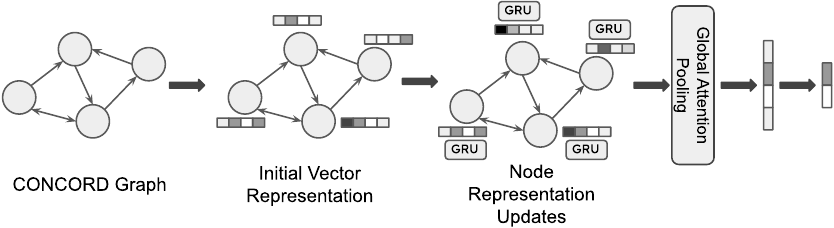}
    \label{fig:ggnn_model}
\end{figure}
Figure~\ref{fig:ggnn_model} illustrates the end-to-end training process of the classifier. 
We take the graph yielded by \concord{} and generate an initial
node feature matrix $V$ using the pre-trained embedding layer of CodeBERT~\cite{feng2020codebert}. 
Once created, we feed $V$ and the adjacency matrix $E$ to the \textsc{ggnn} layer. In this layer, a Gated Recurrent Unit (\textsc{gru})~\cite{Kyunghyun2014} is created for each vertex in $V$. 
The \textsc{ggnn} updates the hidden states of the nodes in an iterative manner, using the following equations:
\begin{eqnarray*}
a_v^t = \sum_{u \in \mathcal{N}(v)} W_{\text{edge}} \cdot h_u^{t-1} \nonumber \\
z_v^t = \text{GRU}_{\text{update}}(a_v^t, h_v^{t-1}) \nonumber \\
h_v^t = (1 - z_v^t) \odot h_v^{t-1} + z_v^t \odot z_v^t \nonumber \\
\end{eqnarray*}
\noindent
where:
\begin{itemize}
    \item $a_v^t$ is the sum of edge weights for node $v$ at time step $t$,
    \item $W_{\text{edge}}$ is a learnable weight matrix for the edges,
    \item $\mathcal{N}(v)$ represents the neighbourhood of node $v$ in the graph,
    \item $\text{GRU}_{\text{update}}(\cdot)$ is the update function of the \textsc{gru},
    \item $z_v^t$ is the update gate for node $v$ at time step $t$,
    \item $h_v^t$ is the hidden state of node $v$ at time step $t$, and
    \item $\odot$ represents element-wise multiplication.
\end{itemize}
After $t$ time steps, the
hidden representations of all the nodes for all the graphs are updated. 
We map the feature matrix $V^{t}$ to a single vector
since we are solving a graph-level classification problem. 
We employ a \textit{Global Attention Pooling} layer~\cite{Li_ggnn2016} that computes an attention score for each node in the graph. 

Finally, we pass the resulting pooled vector into an \textsc{mlp} layer to perform the final classification. 
Table~\ref{tab:hyperparameters} lists all of the hyper-parameters that were used to train the classifier. Our choice of hyperparameters is inspired by the experiments conducted by Siow \etal{}~\cite{Siow2022LearningPS} and Chakraborty \etal{}~\cite{Chakraborty2022}.

\begin{table}[H]
\rowcolors{2}{gray!25}{white}
\centering
\caption{Hyper-parameters and values}
\label{tab:hyperparameters}

\begin{tabular}{lr}
\toprule
Hyper-parameter & Value \\
\midrule
Learning rate & 1e-3 \\
GGNN Layer Input Size & 768 \\
GGNN Layer Hidden Size & 128 \\
MLP Layer Input Size & 128 \\
 MLP Output Size & 2 \\
Batch size & 128 \\
Number of Epochs & 100 \\
Patience & 15 \\
\bottomrule
\end{tabular}
\end{table}

\subsection{Research question}
We train a classifier for each code smell-representation combination, giving a total of 12 trained models. 
Given it is a classification problem, 
we compute  precision ($P$), recall ($R$), and ($F_1$) measure. 
The $F_1$ measure metric is considered insufficient for such classification tasks \cite{Jingxiu2020};
hence, we also include 
Matthews Correlation Coefficient (\textsc{mcc}) metric. 
Furthermore, we compute \textsc{flop}s (floating operation points) to measure the model's complexity. 
The lower the \textsc{flop}s value, the lower the complexity, in turn, the lower the processing.

Through the conducted experiments,
we would like to answer the following research question.

\begin{description}
\item [\textbf{RQ.}] \textit{Does reducing the size complexity of code representation affect the performance and the needed computational operations?}
\end{description}

Answering this research question will highlight the size reduction achieved by \concord{} framework and its effect on
aspects such as computational cost and performance.

\subsection{Results}
Table~\ref{tab:classification_results} shows the performance of the models trained on datasets generated from the three representation configurations. We report the average over five runs with different initialization.
\begin{table}[H]
\centering
\caption{Classification results for each representation. P: Precision, R: Recall, MCC: Matthews Correlation Coefficient, FLOPs: Floating Point Operations.}
\resizebox{1\columnwidth}{!}{%
\begin{tabular}{lccccccccccccccc}
\toprule
\textbf{} & \multicolumn{5}{c}{R1} & \multicolumn{5}{c}{R2} & \multicolumn{5}{c}{R3} \\
\cmidrule(lr){2-6} \cmidrule(lr){7-11} \cmidrule(lr){12-16}
 & P & R & F1 & MCC & FLOPs ($\times$10\textsuperscript{9}) & P & R & F1 & MCC & FLOPs ($\times$10\textsuperscript{9}) & P & R & F1 & MCC & FLOPs ($\times$10\textsuperscript{9})\\
\midrule
\rowcolor{lightgray!25} Complex Method & 0.53 & 0.47 & 0.47 & 0.47 & 0.171 & 0.42 & 0.35 & 0.36 & 0.36 & 0.162 & 0.47 & 0.41 & 0.41 & 0.41 & 0.165 \\
Complex Conditional & 0.48 & 0.43 & 0.44 & 0.44 & 0.175 & 0.43 & 0.40 & 0.40 & 0.40 & 0.168 & 0.47 & 0.45 & 0.45 & 0.45 & 0.170 \\
\rowcolor{lightgray!25} Feature Envy & 0.53 & 0.63 & 0.55 & 0.47 & 0.300 & 0.51 & 0.67 & 0.53 & 0.46 & 0.252 & 0.82 & 0.48 & 0.56 & 0.54 & 0.260\\
Multifaceted Abstraction & 0.52 & 0.16 & 0.24 & 0.25 & 0.232 & 0.52 & 0.13 & 0.21 & 0.21 & 0.205 & 0.52 & 0.13 & 0.21 & 0.23 & 0.210 \\
\bottomrule
\end{tabular}
}
\label{tab:classification_results}
\end{table}
We obtain the average Precision, Recall, F1 measure and MCC as $0.515$, $0.423$, $0.425$, $0.408$ respectively
when training the models using the {\rone} representation.
Although the primary focus of the study is not to showcase the performance of \gnn{}s on code smell detection task, 
one possible explanation for obtaining such values could be the extreme class imbalance of each dataset.
Table~\ref{tab:smell_datasets_stats} shows that the smelly samples accounts for only $\sim4\%$ of the total sample size on average.
Given that we retained the real-life distribution of smelly and benign sample ratio, 
this adds robustness to our evaluation. Indeed, experimenting with datasets that do reflect the actual distribution of anomalies can lead to misleading results. Chakraborty \etal{}~\cite{Chakraborty2022} showed how state-of-the-art \gnn{}-based models used for vulnerability detection failed to yield the same results when trained and evaluated on representative data, and their performance dropped by more than 50\%. 

\begin{tcolorbox}[breakable,width=\linewidth,boxrule=0pt,top=1pt, bottom=1pt, left=1pt,right=1pt, colback=gray!20,colframe=gray!20]
\textbf{Key observation 1:} Graph Neural Network-based models seem to yield a reasonable performance on data that maintains real-world code smell distribution where the presence of smelly snippets is rare. The results indicate that our proposed approach generates code representation that can effectively capture the desired code features for code analysis tasks.
\end{tcolorbox}

Models trained on the \rtwo{} representation show on average $0.47$, $0.38$, $0.37$, and $0.357$ in terms of the classification metrics. 
The models attain around $88\%$ (in terms of \textsc{mcc}) of the original performance compared to the models that were trained using \rone{}.
We performed a paired t-test to measure whether such a difference is significant. We defined the null hypothesis as $H_0$: \textit{The mean difference is zero (\ie{} there is no significant change in {\sc MCC} after removing simple assignment statements)} and the alternative hypothesis as $H_1$: \textit{the mean difference is not zero (\ie{} there is a significant change in MCC after removing such statements)}. We found that the \textit{p-value} = 0.099 $>$ 0.05, hence, we conclude that there is no significant difference in the compared {\sc MCC}s.
In addition, these models exhibit such performance with lower size complexity (\ie{} nodes and edges) graph samples and less computation. 
Table \ref{tab:concord_dataset} shows that graphs under the \rtwo{} representation have  $11.3\%$ fewer nodes and $12\%$ fewer edges on average. 
Consequently, the computational load is reduced by approximately $2.27\times 10\textsuperscript{7}$ \textsc{flop}s.

\begin{tcolorbox}[breakable,width=\linewidth,boxrule=0pt,top=1pt, bottom=1pt, left=1pt,right=1pt, colback=gray!20,colframe=gray!20]
\textbf{Key observation 2:} The elimination of basic assignment statements have resulted in a decrease in the size and complexity of the samples and the associated computational expenses needed for model training. Nonetheless, the models maintained roughly 88\% of their predictive performance, though such a difference is not statistically significant.
\end{tcolorbox}

Interestingly, upon removing print statements as specified in \rthree{}, the trained models  maintain their performance on par with the models trained using \rone{} representation (in terms of \mcc{}).
The removal of print statements also resulted in an increase of \mcc{} scores for the \textit{Complex Conditional} and \textit{Feature Envy} code smells by $2.2\%$ and $14.9\%$, respectively. 
Additionally, the models achieved comparable results for the \textit{Complex Method} and \textit{Multifaceted Abstraction} code smells, with \mcc{} scores of $0.41$ and $0.47$, and $0.23$ and $0.25$, respectively. 
Similar to the \rtwo{} representation, 
the \rthree{} graphs are smaller in size, 
having $8\%$ fewer nodes and $9\%$ fewer edges compared to \rone{}. 
Therefore, graphs generated from \rthree{} require less computation, with the reduction of floating point operations by approximately $1.87 \times 10\textsuperscript{7}$ operations.

\begin{tcolorbox}[breakable,width=\linewidth,boxrule=0pt,top=1pt, bottom=1pt, left=1pt,right=1pt, colback=gray!20,colframe=gray!20]
\textbf{Key observation 3:} Adopting \texttt{print} statement removal as a pruning technique has enabled the models to retain their predictive performance at a rate of 100\%. In addition to maintaining this high-performance rate, the approach has also resulted in reduced graph sample sizes, leading to lower computational costs.
\end{tcolorbox}

Both strategies \ie{} simple assignment removal (in \rtwo{}) 
and print statement removal (in \rthree{}) 
reduce computational costs by decreasing the number of \textsc{flop}s performed. 
The \rtwo{} representation requires approximately $\times 1.02$ fewer operations than \rthree{}. 
However, in terms of performance, \rthree{} is superior
as illustrated in the results. 
It manages to maintain the same level of performance as the models trained on the original data, 
whereas \rtwo{} did not fully attain the performance of {\rone}. 
The choice between these strategies depends on a specific goal. 
If the objective is to reduce computation, 
adopting the simple assignment strategy could be a suitable option, 
as it can, to some extent, maintain the original performance. 
However, if the goal is to balance both performance and computation, removing print statements might be a better option. 
This strategy reduces computation while fully preserving performance.

\begin{tcolorbox}[breakable,width=\linewidth,boxrule=0pt,top=1pt, bottom=1pt, left=1pt,right=1pt, colback=gray!20,colframe=gray!20]
\textbf{Key observation 4:} Both pruning strategies have led to a decrease in computational costs given they have reduced the input size of the \gnn{}-based models. However, removing print statements seems to be more a efficient strategy in this context because it succeeded in fully retaining the performance and reducing the complexity of graphs at the same time.
\end{tcolorbox}

In order to further investigate the impact of removing print statements, we select samples from the test set of the \textit{Complex Conditional} dataset where the models trained on {\rone} and \rthree{} exhibit disagreement by providing different classifications. 
Listing~\ref{lst:noisy_prints_snippet} presents a snippet where such a disagreement occurred. 
For presentation purposes, 
the code has been simplified, 
whereas the full version can be found here\footnote{\url{http://bit.ly/446v0n7}}.



\begin{listing}[ht]
\begin{minted}[
fontsize=\small,
escapeinside=||
]{java}
public TwoPassDataIndexer(...PARAMS) {
    TObjectIntHashMap predicateIndex;
    List eventsToCompare;
    predicateIndex = new TObjectIntHashMap();
    |\colorbox{yellow}{System.out.println(STRING);}|
    |\colorbox{yellow}{System.out.print(STRING);}|
    try {
      File tmp = File.createTempFile("events", null);
      tmp.deleteOnExit();
      int numEvents = ...;
    |\colorbox{yellow}{System.out.println(STRING);}|
    |\colorbox{yellow}{System.out.print(STRING);}|
      eventsToCompare = ...;
      // done with predicates
      predicateIndex = null;
      tmp.delete();
    |\colorbox{yellow}{System.out.println(STRING);}|
    |\colorbox{yellow}{System.out.print(STRING);}|
      sortAndMerge(eventsToCompare);
    |\colorbox{yellow}{System.out.println(STRING);}|
    }
    catch(IOException e) {
    |\colorbox{yellow}{System.out.println(e);}|
    }
\end{minted}
\caption{An example showing noisy print statements affecting the predictive performance. It is a benign sample from \textit{Complex conditional} smell ground truth. Models trained on \rone{} representation detects the smell whereas models trained on \rthree{} classifies it as benign.}
\label{lst:noisy_prints_snippet}
\end{listing}

The model trained on {\rone} classifies the code snippet as smelly, whereas the model trained on \rthree{} classifies it correctly. 
Misclassification can be attributed to the \textit{noise} that print statements introduce, as they lead the model to learn implicit patterns that potentially hinder the learning of an effective representation.
Analogous to stop words in the Natural Language Processing (NLP) domain, print statements, in this context, 
introduce noise and do not help the model to classify correctly. 
Moreover, their high frequency increases computational costs.

\concord{} \dsl{} and framework automates the customized code representation.
Of course, the automation and simplification strategies offered by \dsl{} must be used according to the requirements of the downstream task. For example, it may be unwise to perform print statement removal if we would like to generate code representation for code search as the downstream task that retrieves snippets that print to the I/O. Continuing with the analogy we drew from NLP, the same can be applied to stop words. For instance, if an NLP system is used for stylistic analysis (\eg{} authorship attribution) removing stop words may remove important stylistic markers, such as sentence structure and word choice.
\section{Threats to Validity}
\topic{Internal validity}
Internal validity threats are related to experiment errors. When we cloned the repositories, we took into account the snapshots that were prior to the publication date of the replication package released by Sharma \etal{}~\cite{Sharma2021A}. This was done to minimize the risk of code smells being refactored. Moreover, we excluded code snippets that did not have simple assignment and print statements to avoid generating misleading results. Furthermore, during the sampling process for the construction of the data set, we ensured the preservation of the same positive-to-negative class ratio to avoid erroneous results as discovered by Chakraborty \etal{}~\cite{Chakraborty2022}. Furthermore, to ensure fairness in evaluation, all models were trained and evaluated on the exact same splits.

\topic{External validity}
This type of threat is related to the generalizability of the method. Although we provided a use case for code smell detection, \concord{} can be integrated into many other tasks, such as vulnerability detection or code search. Moreover, it supports multiple programming languages, unlike the aforementioned tools, which makes it flexible and can be used on multi-linguistic datasets and benchmarks.

\topic{Construct validity}
Construct validity threats are related to measurements and randomness. First, we carefully constructed the {\rone}, {\rtwo}, and {\rthree} configurations to measure the effect of removing simple assignment and print statements across the code snippets in isolation. Moreover, we set the same random seed across all models during the training and inference phases. Regarding metrics, we used adequate measures to evaluate classifiers and calculate their computational costs as done in similar work~\cite{Sharma2021A, Zhang2022, Rabin2021, Kim2022}.
\section{Conclusions, Implications, and Future Work}
In this study, we propose \concord{}, a \dsl{} for graph code representation construction. It is a method to build representations in a configurable and customizable way.
We demonstrate how it can be used in a real example to detect code smells. Our experiments show that it can effectively produce graph representation of source code and reduce computational costs while maintaining up to 100\% performance.

\textit{Implications for Research:} Using this \dsl{}, researchers no longer need to manually implement disparate edge augmentations and modelling configurations, saving time and effort. Moreover, by introducing a unified tooling support, researchers can adopt a standardized approach to code representation and analysis, enhancing the reproducibility of experiments and fostering transparency in research findings. Having a tool like \concord{} that unifies the fragmented functionality of graph-based code representations 
and allows composition can accelerate research progress in the field. Researchers can quickly experiment with different configurations, facilitating faster iteration and hypothesis testing. The ease of use and automation provided by the tooling support can lead to more efficient exploration of new analysis techniques and advancing the state-of-the-art in software engineering research. We also hope that by open-sourcing our work, the \dsl{} can be further expanded by the research community to accommodate new emerging ideas.

\textit{Implications for Industry:} In the industry context, \concord{} addresses the challenge of scalability in training \gnn{}s. The reduction heuristics can contribute to the creation of scalable solutions for bug detection, code refactoring, and security analysis in industries. In addition, the reduced engineering overhead of using \concord{} facilitates quicker adoption of graph-based source code analysis in real-world software development pipelines, potentially improving code quality and software maintenance practices.

In the future, we will investigate how we can apply \concord{} in other tasks in source code analysis, and extend it further to accommodate other modelling techniques that can emerge in the literature. We hope that this \dsl{} helps the research community explore new frontiers by providing new tooling support.



\bibliographystyle{ACM-Reference-Format}
\bibliography{refs}

\end{document}